\documentclass[conference,compsoc]{IEEEtran}
\ifCLASSOPTIONcompsoc
  \usepackage[nocompress]{cite}
\else
  \usepackage{cite}
\fi
\usepackage{epsfig,endnotes}
\usepackage{tabularx}
\usepackage{adjustbox}
\usepackage{multirow}
\usepackage{makecell}
\usepackage{amsfonts}
\usepackage{amsmath} 
\usepackage{hhline}
\usepackage{enumitem}
\usepackage{booktabs}
\usepackage{caption}
\usepackage{seqsplit}
\usepackage{hyperref}
\usepackage{xcolor}
\usepackage{tikz}

\def\ourmethod{ByzSFL}
\def\SAframework{DuoAgg}

\newcommand{\para}[1]{\vspace{2pt}\noindent\textbf{{#1}}}
\newcommand{\ignore}[1]{}
\newcommand*\circled[1]{\tikz[baseline=(char.base)]{
            \node[shape=circle,draw,inner sep=2pt] (char) {#1};}}

\hypersetup{
  colorlinks,
  linkcolor={blue!70!green},
  citecolor={green!70!blue},
  urlcolor={orange!70!red}
}

\ifCLASSINFOpdf
\else
\fi
\begin{document}
%
\title{ByzSFL: Achieving Byzantine-Robust Secure Federated Learning with Zero-Knowledge Proofs}

\author{
\IEEEauthorblockN{
Yongming Fan\IEEEauthorrefmark{1}\IEEEauthorrefmark{4}, 
Rui Zhu\IEEEauthorrefmark{1}\IEEEauthorrefmark{3}, 
Zihao Wang\IEEEauthorrefmark{2}, 
Chenghong Wang\IEEEauthorrefmark{2}, 
Haixu Tang\IEEEauthorrefmark{2}, \\
Ye Dong\IEEEauthorrefmark{5},
Hyunghoon Cho\IEEEauthorrefmark{3}, 
Lucila Ohno-Machado\IEEEauthorrefmark{3}
}\\
\IEEEauthorblockA{\IEEEauthorrefmark{2}Indiana University Bloomington, Bloomington, IN, USA \\
Email: \{zwa2, cw166, hatang\}@iu.edu}
\IEEEauthorblockA{\IEEEauthorrefmark{3}Yale University, New Haven, CT, USA \\
Email: \{rui.zhu.rz399, hoon.cho, lucila.ohno-machado\}@yale.edu}
\IEEEauthorblockA{\IEEEauthorrefmark{4}Purdue University, West Lafayette, IN, USA \\
Email: fan322@purdue.edu}
\IEEEauthorblockA{\IEEEauthorrefmark{5}Singapore University of Technology and Design, Singapore\\
Email: ye\_dong@sutd.edu.sg}
\IEEEauthorblockA{\IEEEauthorrefmark{1}These authors contributed equally to this work.}
}
\maketitle

\begin{abstract}

The advancement of AI models, especially those powered by deep learning, faces significant challenges in data-sensitive industries like healthcare and finance due to the distributed and private nature of data. Federated Learning (FL) and Secure Federated Learning (SFL) enable collaborative model training without data sharing, enhancing privacy by encrypting shared intermediate results. However, SFL currently lacks effective Byzantine robustness—a critical property that ensures model performance remains intact even when some participants act maliciously. Existing Byzantine-robust methods in FL are incompatible with SFL due to the inefficiency and limitations of encryption operations in handling complex aggregation calculations. This creates a significant gap in secure and robust model training.

To address this gap, we propose \ourmethod{}, a novel SFL system that achieves Byzantine-robust secure aggregation with high efficiency. Our approach offloads aggregation weight calculations to individual parties and introduces a practical zero-knowledge proof (ZKP) protocol toolkit. This toolkit supports widely used operators for calculating aggregation weights, ensuring correct computations without compromising data privacy. Not only does this method maintain aggregation integrity, but it also significantly boosts computational efficiency, making \ourmethod{} approximately 100 times faster than existing solutions. Furthermore, our method aligns with open-source AI trends, enabling plaintext publication of final model without additional information leakage, thereby enhancing the practicality and robustness of SFL in real-world applications.

\end{abstract}


%
\IEEEpeerreviewmaketitle

\section{Introduction}

Over the past decade, the advancement of AI, particularly driven by deep learning and neural networks, has heavily relied on the availability of increasingly large datasets. From AlphaGo to ChatGPT, these data-hungry deep learning models require vast amounts of data to ensure their effectiveness. However, in many critical sectors such as healthcare and finance, data is often distributed across multiple parties (also known as clients). In these sectors, the data is highly sensitive, while each party may be unwilling or legally unable to share their data directly. This limitation means that each party can only develop AI models using its own data, which often lacks the scale needed for effective AI development. Consequently, the progress of AI in these sectors has lagged behind comparing with other sectors.

To address this challenge, techniques such as \textit{federated learning (FL)} and collaborative learning were developed. These methods allow parties to train models collectively without directly sharing their data with a central server, instead only sharing specific intermediate results, thus preserving privacy. Building on this foundation, \textit{secure federated learning (SFL)} was later introduced, which further enhances security by encrypting the shared intermediate results, thereby protecting against potential privacy leaks. This advancement has significantly propelled AI development in data-sensitive industries.

Specifically, the FL framework typically involves each party following predefined computation rules to calculate intermediate results based on their local datasets and the current state of the global model. These intermediate results are then uploaded to a central server, which aggregates them according to certain rules. A common aggregation method assigns weights based on the amount of data each party contributes, giving more influence to parties with larger datasets. However, a significant challenge within FL is that the final model's performance can be disproportionately affected by the intermediate results of just a few parties. This issue may arise due to several reasons: the data from these parties might differ significantly from that of others, leading to distributional skewness; the local computations might be prone to errors; or, more concerningly, the parties might act maliciously. For instance, a party might aim to skew the final model to better fit its own data or even launch a poisoning attack.

To address these challenges, the concept of \textit{Byzantine robustness} was introduced in FL. A Byzantine-robust FL model can maintain its accuracy and integrity even when a minority of the parties are malicious, regardless of what these parties upload. Numerous approaches have been developed to achieve Byzantine robustness~\cite{fltrust, Trim-mean, Krum}, aiming to mitigate the influence of adversarial participants. These methods typically apply operations such as norm~\cite{Krum}, absolute value~\cite{fedavg}, median~\cite{Median}, mean~\cite{Trim-mean}, or cosine similarity~\cite{fltrust} to the uploaded intermediate results, and then recompute these results. Based on the outcomes, these approaches either assign different aggregation weights or decide whether to include the results in the aggregation process.


However, achieving Byzantine robustness in SFL poses a dilemma. There are two main types of homomorphic encryption: \textit{Partially Homomorphic Encryption (PHE)} and \textit{Fully Homomorphic Encryption (FHE)}. PHE is significantly more efficient than FHE, often by a factor of 85 or more~\cite{sarkar2021fast, cominetti2020fast}, but it is limited to basic operations, such as addition and subtraction on encrypted data. Consequently, when using PHE, the server can only perform simple aggregations, such as addition, while all other computations must be offloaded to the clients. This reliance on clients for aggregation weight calculations introduces vulnerabilities, as careless or malicious clients can easily exploit these weaknesses.

On the other hand, SFL systems using FHE allow the central server to handle complex computations, including aggregation weight calculations, thus mitigating potential security loopholes. However, this comes at a significant cost to efficiency, making the overall system much slower. As a result, current SFL implementations are often limited to simple norm-based input validation techniques~\cite{ACORN, hao2021efficient, RoFL}, which only check the norm of encrypted intermediate results. Such simplistic validation methods fall short of achieving true Byzantine robustness, as state-of-the-art robustness mechanisms often require more sophisticated operations, such as cosine similarity~\cite{fltrust} or robust averaging methods~\cite{Trim-mean}. Consequently, a substantial gap remains in developing SFL systems that are both secure and efficient.

To address this gap, we propose \ourmethod{}, an innovative SFL system that maintains Byzantine robustness without incurring significant overhead. \ourmethod{} leverages \textit{Partially Homomorphic Encryption (PHE)} to encrypt the intermediate results shared by the parties, while offloading the calculation of aggregation weights to individual clients. Each client computes its own aggregation weight locally, which introduces the risk of incorrect weights being submitted, either accidentally or maliciously. To mitigate this risk, \ourmethod{} incorporates a Zero-Knowledge Proof (ZKP) protocol. This protocol enables clients to demonstrate—without revealing any sensitive information—that their aggregation weight was accurately derived from their encrypted intermediate results using the FLTrust score. By ensuring the integrity of aggregation weights and maintaining efficiency, \ourmethod{} effectively bridges the gap in secure and robust SFL systems.


\para{Contributions}.
Our key contributions are outlined below:

\vspace{2pt}\noindent$\bullet$ We propose \ourmethod{}, the first system to achieve highly efficient Byzantine-Robust Secure Aggregation in SFL. \ourmethod{} ensures robust model training even in the presence of malicious participants, addressing a critical gap in existing SFL frameworks.

\vspace{2pt}\noindent$\bullet$ We design a ZKP protocol that ensures correct aggregation weights without revealing sensitive data, achieving a 100x speedup over existing Byzantine-robust FL methods in SFL.
    
\vspace{2pt}\noindent$\bullet$ 
We present \SAframework{}, a dual-server design that strengthens security and efficiency under practical assumptions, surpassing traditional masking and homomorphic encryption-based frameworks.


\section{Background}
\label{sec:Background}

\subsection{FL and Secure Aggregation for FL}
\label{subsec:FL and Secure Aggregation for FL}

Federated Learning (FL) is a distributed framework~\cite{zhu2020privacy,zhang2021survey,xu2021federated} in which multiple clients collaboratively learn a model under the coordination of a central server. In each round of FL, first the server broadcasts the current model to all clients participating in that round. Then, second step, a local training step, clients update the model using their own local datasets. Finally, the third step, the clients proceed to an aggregation step, where all locally trained models are pre-processed and aggregated by the server to obtain an updated model. These steps (1-3) are repeated over several training rounds.

The critical role of secure aggregation occurs in step three, which distinguishes secure Federated Learning (SFL) from standard Federated Learning. In this step, secure aggregation is employed to significantly reduce the data leakage that would otherwise occur if model updates were directly exposed to the server. This is achieved by encrypting the locally updated model vectors that clients send to the central server, which then aggregates these encrypted vectors. For a more detailed discussion on the typical processes involved in secure aggregation, we refer to Section~\autoref{sec:Generalized Secure Aggregation Framework}.

\subsection{Input Validation \& Byzantine-robust}
\label{subsec:Input Validation}

When considering non-secure aggregation (without encryption), the large number of clients and the complexity of data exchange processes often introduce significant uncertainty in the final model’s performance. These uncertainties can arise due to inadvertent mistakes or adversarial intentional actions, such as poisoning attacks, by individual clients during the model update process. To mitigate these issues, it is common to calculate an \textit{aggregation weight} for each client at the server to manage such risks. For example, researchers initially proposed simple constraints on the vectors uploaded by clients, such as norm bounds, where vectors with larger norms would be assigned smaller aggregation weights or might even be discarded. However, these measures have proven only moderately effective~\cite{fltrust}. Subsequently, more sophisticated input validation methods have been introduced in FL, aiming to achieve \textit{Byzantine-robustness}. In essence, Byzantine-robust Federated Learning enables a service provider to learn an accurate global model even when a bounded number of clients are malicious. Below, we introduce the existing methods for Input Validation in both non-secure and secure aggregation processes.

\para{Input Validation in Aggregation}.
Input Validation in Aggregation involves the implementation of techniques designed to ensure that the model updates provided by clients are both reliable and secure, thereby maintaining the performance and integrity of the global model. This is particularly important in the presence of Byzantine faults, where some clients may act maliciously or submit incorrect updates, either due to errors or deliberate intent.

Byzantine-robustness~\cite{DBLP:conf/nips/BlanchardMGS17, DBLP:conf/sigmetrics/ChenSX18, DBLP:conf/icml/YinCRB18} in this context refers to the ability of the FL system to withstand and operate effectively despite the presence of such malicious clients. The challenge lies in ensuring that the global model remains accurate and unbiased, even when a portion of the participating clients is actively attempting to compromise the training process. Various strategies have been developed to address this challenge, focusing on filtering out or mitigating the impact of potentially harmful updates.

One of the straightforward forms of input validation is norm bounding, which involves setting limits on the norms of the model updates sent by clients. This approach restricts the influence of any single client on the global model, thereby reducing the risk of the model being skewed by outlier updates. However, while norm bounding is straightforward, it may not be sufficient in scenarios with more sophisticated attacks.

To address more complex threats, methods like Krum~\cite{Krum} have been proposed. Krum is an aggregation rule that evaluates the squared distance between a client’s model update and the updates from other clients. Specifically, Krum computes a score for each client by summing the distances between its update and those of the closest $n-f-2$ clients, where $f$ represents the number of potentially malicious clients. The client with the minimal score is then selected to update the global model. This method is particularly effective when the number of malicious clients is small, as it minimizes the impact of outlier updates by focusing on the consensus among the majority of clients.

Another approach is the Trimmed Mean (Trim-mean)~\cite{Trim-mean} method, which aggregates model updates in a coordinate-wise manner. For each parameter in the model, the server collects all values from the clients, sorts them, and then removes the largest and smallest values according to a trimming parameter $k$. The mean of the remaining values is used to update the global model. Trim-mean can tolerate up to 50\% of the clients being malicious, making it a robust method in scenarios with a higher proportion of adversarial clients.

Similarly, the Median method~\cite{Median} operates by sorting the values of each model parameter across all client updates. Instead of averaging the trimmed values as in Trim-mean, the median value is selected for the global model. This method is particularly simple but effective against outliers and certain types of malicious behavior, offering a balance between simplicity and robustness.

Finally, FLTrust~\cite{fltrust} is widely recognized as one of the most advanced approaches due to its incorporation of a trust bootstrapping mechanism. In FLTrust, the server utilizes a small, trusted dataset to compute a reference upddate for each round. It then evaluates the similarity between the updates from clients and this reference update using a variant of cosine similarity, known as the \textit{FLTrust score}. The FLTrust score is defined as $\text{FLTrust score} = \max(0, \text{cossim}(v_{client}, v_{validate}))$, where \textit{cossim} denotes cosine similarity, $v_{client}$ represents the local model updates from client, and $v_{validate}$ represents the reference update computed by the server.

The FLTrust score is then used as an aggregation weight to adjust the influence of each client. By implementing this mechanism, FLTrust not only validates the inputs but also significantly enhances the overall robustness of the aggregation process, providing a stronger defense against Byzantine failures.

\para{Input Validation in Secure Aggregation}.
In Secure Aggregation, the challenge of input validation is compounded by the encryption of model updates, which complicates the application of traditional validation methods like norm bounding. To tackle these challenges, recent research has focused on ensuring that the norms of encrypted vectors remain within acceptable limits. While such approaches offer a basic level of security, they often suffer from significant computational and communication overheads, limiting their practicality in large-scale deployments.

One of the earliest and still prevalent approaches is zPROBE by Ghodsi et al.~\cite{zPROBE}, which employs a generic proof framework to verify that each entry of a client’s masked input is honestly derived from an input of bounded size. However, due to the high computational cost, this method only verifies a random subset of entries. This selective verification might not be robust enough for Federated Learning, where even small deviations in a few entries could compromise the entire model.

Similarly, Karakoç et al.~\cite{karakocc2021secure} have proposed a method involving secure aggregation with range validation through an oblivious programmable pseudorandom function. However, their work remains at the proof-of-concept stage, with experimental validation limited to very small vector lengths (e.g., 16), due to the prohibitive computational costs involved.

Methods like RoFL\cite{RoFL} and EIFFeL\cite{EIFFeL}, introduced by Lycklama et al. and Chowdhury et al. respectively, have focused on more complex setups. RoFL, for instance, requires each client to send commitments for every vector entry, leading to substantial communication overhead. EIFFeL distributes the computational load across clients, but faces scalability issues, with communication overheads increasing quadratically with the number of clients and linearly with the vector length. The challenge of balancing input validation with communication costs continues to be a major focus in the development of Secure Aggregation protocols.

\subsection{Homomorphic Encryption}
\label{subsec:Homomorphic Encryption}
Homomorphic encryption is a class of cryptographic algorithms that allow computations to be performed directly on encrypted data without needing to decrypt it first~\cite{acar2018survey}. This means that the computing party, such as a cloud service provider, can process data while it remains encrypted, ensuring that sensitive information is never exposed in its unencrypted form. This capability is crucial in scenarios where privacy and security are paramount, such as in financial services~\cite{peng2016homomorphic}, healthcare~\cite{munjal2023systematic}, and data analytics~\cite{mr2016homomorphic}.

Partial homomorphic encryption (PHE) and Fully homomorphic encryption (FHE)~\cite{zhang2016review} are two major types of homomorphic encryption. Partial homomorphic encryption (PHE) permits only a single type of computation on encrypted data, either addition or multiplication. Fully homomorphic encryption (FHE), on the other hand, enables arbitrary computations on encrypted data, offering a powerful solution for secure data processing. However, FHE requires considerable computational resources, making it more demanding and slower than PHE. The following sections outline the latest methods in both PHE and FHE, with a focus on the specific PHE technique used in our approach and the FHE method chosen for comparison.

\para{PHE}.
The Paillier cryptosystem is a partially homomorphic encryption scheme that supports homomorphic addition of ciphertexts, providing both efficiency and strong security. This scheme operates over a composite modulus \( n = pq \), where \( p \) and \( q \) are large prime numbers. To encrypt a message \( m \), the ciphertext \( c \) is computed as:
\[
c = g^m \cdot r^n \mod n^2
\]
where \( g \) is an element of high order in \( \mathbb{Z}_{n^2}^* \), and \( r \) is a random value selected for each encryption. The Paillier scheme's homomorphic property allows the multiplication of two ciphertexts \( c_1 \) and \( c_2 \) to produce a new ciphertext \( c_3 = c_1 \cdot c_2 \mod n^2 \), which decrypts to the sum of the original plaintexts, i.e., \( m_1 + m_2 \). Although it does not enable homomorphic multiplication, Paillier encryption is highly suitable for applications requiring additive operations on encrypted data, such as secure voting and privacy-preserving data aggregation. The composite modulus \( n \) enhances security against known attacks, making the Paillier cryptosystem an effective choice for contexts that require efficient homomorphic addition.

\para{FHE}.
The CKKS (Cheon-Kim-Kim-Song) protocol is a homomorphic encryption scheme tailored for performing approximate arithmetic on encrypted data, making it highly suitable for privacy-preserving computations on real or complex numbers~\cite{cheon2017homomorphic}. In this scheme, a plaintext value \( m \) is first encoded into a polynomial \( m(x) \) within the ring \( \mathbb{R}[x]/(x^n + 1) \), where \( n \) is the polynomial modulus degree. The encoded polynomial is then scaled by a factor \( \Delta \), which controls the precision of subsequent arithmetic operations. The encryption of the plaintext is expressed as:
\[
c(x) = a(x) + \Delta \cdot m(x) + e(x) \pmod{q(x)}
\]
where \( a(x) \) is a random polynomial, \( e(x) \) is an error polynomial introduced during encryption, and \( q(x) \) is the ciphertext modulus. The noise \( e(x) \), which accumulates with each homomorphic operation, necessitates careful parameter selection to balance precision, efficiency, and security within the protocol.



\subsection{Zero-Knowledge Proof and zk-SNARKs}
Zero-Knowledge Proof (ZKP) ~\cite{goldreich2019proofs} is a fundamental concept in cryptography that enable one party, known as the prover, to demonstrate to another party, the verifier, that they possess a certain piece of information or can perform a specific computation without revealing the actual information or details of the computation itself. This unique capability has profound implications for enhancing privacy and security in various applications, ranging from authentication systems to secure multiparty computations. The core idea behind ZKPs is to provide assurance that the prover is truthful without requiring them to disclose sensitive data, thereby preventing any potential leakage of information during the verification process.

\para{zk-SNARKs.} Zero-Knowledge Succinct Non-Interactive Arguments of Knowledge (zk-SNARKs) ~\cite{ben2013snarks, parno2016pinocchio, groth2016size} enhance the privacy-preserving features of Zero-Knowledge Proofs (ZKPs) by incorporating additional attributes such as succinctness and non-interactivity. These features enable zk-SNARKs to produce highly compact proofs that require minimal communication between the prover and verifier, thereby significantly improving efficiency and scalability. Succinctness ensures that the proofs remain brief, even for complex computations, while non-interactivity eliminates the need for multiple rounds of communication between the parties, making the protocol practical for real-world applications, including those in machine learning and artificial intelligence~\cite{han2023kick, zhu2023robust}. 

At a high level, the first step in zk-SNARKs involves transforming the statement to be proven into an equivalent form that hinges on solving algebraic equations represented as an arithmetic circuit. To verify the correct evaluation of this circuit, the developer must impose a series of constraints on the circuit's wires, known as a constraint system, which is typically a Rank-1 Constraint System (R1CS). Converting the proving statement into an R1CS circuit often requires manual conversion, which can introduce security vulnerabilities. However, there are now applications designed to provide end-to-end security evaluations for zk-SNARK proofs and programs, ensuring robust security~\cite{fan2024snarkprobe, pailoor2023automated}. Additionally, tools like Circom, a compiler that assists developers in constructing R1CS, have been developed to reduce the difficulty associated with deploying zk-SNARK programs. Several tools \cite{fan2024snarkprobe} \cite{pailoor2023automated} have been developed to assess the correctness and security of zk-SNARK proofs.

\section{Preliminary}
\label{sec:Preliminary}

\subsection{Motivation and Challenge}
\label{subsec:Motivation and Challenge}
Byzantine-robust federated learning aims to enable a service provider to train an accurate global model even when some clients behave maliciously. This robustness is crucial in addressing the uncertainties inherent in federated learning, where numerous parties are involved, and the computation is complex and distributed. However, no existing method achieves Byzantine robustness in the context of secure aggregation within federated learning. Achieving Byzantine-robust secure aggregation is particularly challenging because it requires validating encrypted vectors, significantly complicating the process.

Current research on secure aggregation has only implemented basic input validation methods, such as verifying whether the norm of an encrypted vector lies within a predefined range~\cite{ACORN}. While seemingly simple, even these basic methods require highly complex protocols~\cite{ACORN}. Consequently, advanced input validation techniques, such as FLTrust~\cite{fltrust}, which are effective in standard Byzantine-robust federated learning, cannot be directly applied in the secure aggregation context.

To address this gap, we propose \ourmethod{}, an efficient Byzantine-robust secure aggregation framework for federated learning. Our approach offloads the computation of aggregation weights to clients, significantly reducing the computational burden on the central server. This design enables the server to focus solely on aggregating encrypted data, leveraging more efficient Partially Homomorphic Encryption (PHE).

However, this shift introduces a challenge: aggregation weights computed by clients may be incorrect due to errors or malicious behavior. To mitigate this, \ourmethod{} incorporates a comprehensive Zero-Knowledge Proof (ZKP) protocol toolkit. This toolkit enables clients to prove that their aggregation weight computations are valid without revealing sensitive information. We design proofs for a wide range of operators, making the toolkit compatible with most methods for calculating aggregation weights in the literature~\cite{Krum, Trim-mean, fedavg, Median}. This innovation fulfills the Byzantine robustness requirements for secure federated learning.

\subsection{Threat Model}
\label{subsec:Threat Model}

We consider a scenario involving $m$ clients collaborating to develop a model in a privacy-preserving manner. Initially, a key generator distributes a shared public key and the initial model to all clients. Each client uses its local data and the initial model to compute an intermediate result (e.g., gradients) based on publicly shared training hyperparameters. The intermediate result is then encrypted using the public key. A central server, denoted as $\mathbb{S}_C$, receives these encrypted intermediate results and performs the aggregation. After completing the aggregation, $\mathbb{S}_E$ receives the encrypted aggregated result from $\mathbb{S}_C$, computes the updated encrypted model, and sends it back to the clients, completing one round of the training process.

\para{Client Goals and Capabilities}.  
We assume the majority of clients act honestly, following instructions to contribute to the final model. Clients control the encrypted vectors they send to $\mathbb{S}_C$, the aggregation weights they use, and the generation of ZKP proofs. However, some clients may act dishonestly, modifying their data or altering aggregation weights either accidentally or deliberately to bias the final model in their favor. We assume only a subset of clients may behave dishonestly, as effective training would be infeasible if most or all clients were dishonest. The goal of our method is to leverage ZKP to identify and mitigate the influence of such clients.


\para{$\mathbb{S}_E$ and $\mathbb{S}_C$ Assumptions}.  
We assume that $\mathbb{S}_E$ has access to a small subset of data ($D^*$) that is representative of the target distribution for model training. This subset is intentionally designed to have low privacy sensitivity, as it may consist of publicly available data or a demo dataset. This setup aligns with the configuration used in FLTrust~\cite{fltrust}. There is no collusion between $\mathbb{S}_E$ and $\mathbb{S}_C$, or between $\mathbb{S}_E$ and any client. 
Since $\mathbb{S}_C$ has access only to encrypted data and does not possess the decryption key, it does not pose a significant threat. However, $\mathbb{S}_E$ learns the aggregated result at each model update phase. While this is consistent with the threat models considered in existing SFL studies~\cite{RoFL, EIFFeL, ACORN}, we provide a detailed analysis of potential security and privacy risks associated with malicious behavior by $\mathbb{S}_E$ in \autoref{subsec:Privacy and Security Analysis}.

\subsection{Overall Setup}


Considering the setup described above and using a neural network model $f_{\beta}$ as an example, we focus on secure aggregation. The initial model parameters are denoted as $\beta_0$. The server $\mathbb{S}_E$ uses its validation dataset $D^* = \{ X^*, Y^* \}_{n^*}$ and a given loss function. Without loss of generality, we consider the mean square error (MSE) as our loss function for a regression task, defined as:

\begin{equation}
\label{eq:ell}
\ell (\beta;D^*)= \frac{1}{n^*} \sum_{j=1}^{n^*} \left( Y_j^* - f_{\beta}(X_j^*, \beta)\right)^2
\end{equation}

The server $\mathbb{S}_E$ computes a reference update by minimizing the loss function $\ell$ using a specified optimizer, $\mathcal{O}$ (such as Stochastic Gradient Descent or similar methods), with a local learning rate $\eta$. This reference update is defined as:

\begin{equation}
\label{eq:beta^*}
\bold{g}^* = \eta \cdot \nabla_{\beta} \ell(\beta;D^*)
\end{equation}

Here, $\bold{g}^*$ represents the gradient update that minimizes the loss function $\ell$ on the validation dataset $D^*$.

For the $i^{th}$ client, with its dataset $D^{(i)} = \{ X^{(i)}, Y^{(i)} \}_{n^{i}}$, the corresponding update is denoted as $\bold{g}_i$ and is defined as:
\begin{equation}
\label{eq:beta_i}
\bold{g}_i = \eta \cdot \nabla_{\beta} \ell(\beta;D^{(i)})
\end{equation}

Note that, in this work, we adopt the state-of-the-art Byzantine robustness approach, FLTrust~\cite{fltrust}, which employs a sophisticated operator (cosine similarity) to enhance model reliability.

The Trust Score (TS) from FLTrust~\cite{fltrust}, which serves as the unnormalized aggregation weight for the $i^{th}$ client, is defined as:
 \begin{equation}
 \label{eq:fltrust score}
 \text{TS}_i = \max(0, \frac{\bold{g}^* \cdot \bold{g}_i}{\|\bold{g}^*\| \|\bold{g}_i\|})
 \end{equation}

The normalized trust score denoted as $\tilde{\text{TS}}_i$:

\begin{equation}
\label{eq:normalizd fltrust score}
\tilde{\text{TS}}_i =  \text{TS}_i \cdot \frac{\left\|\bold{g}^*\right\|}{\left\|\bold{g}_i\right\|}
\end{equation}
  
The $i^{th}$ client computes its encrypted weighted local model, and denote as:
\begin{equation} 
\label{eq:client upload}
 \mathcal{E} ( \tilde{\text{TS}}_i \cdot \bold{g}_i)
\end{equation} 
 


For the cryptographic aspects, we use \texttt{ek} as the encryption (public) key and \texttt{dk} as the decryption (private) key for homomorphic encryption. $\mathcal{E}_{\texttt{ek}}$ and $\mathcal{D}_{\texttt{dk}}$ represent Encrypt and Decrypt operations, respectively. 
For zero-knowledge proofs, \texttt{pk} is used as the proving key and \texttt{vk} as the verification key. Encrypted variables are denoted using $\mathcal{E}$.
We have summarized these commonly used notations in Appendix~\ref{sec:Notation}, \autoref{tab:notations}.


\section{\SAframework{}: A Generalized Secure Aggregation Framework}
\label{sec:Generalized Secure Aggregation Framework}

\begin{figure}[!t]
\centerline{\includegraphics[width=\linewidth]{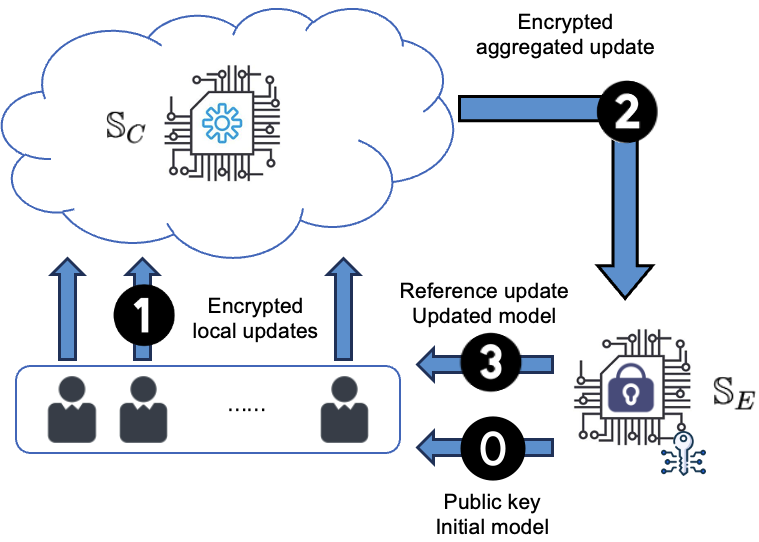}}
\caption{\small Framework of ~\SAframework {}}
\vspace{-15pt}
\label{fig:fig_1}
\end{figure}


In this chapter, we focus on explaining the framework of \ourmethod{} without the input validation components. This allows us to dedicate the next chapter, \autoref{sec:ByzSFL}, to specifically introducing the design of our input validation protocol. We name this general framework as \textit{\SAframework{}}. A schematic representation of the architecture is provided in \autoref{fig:fig_1}.

To enhance the efficiency of our protocol, we developed \SAframework{} by distributing responsibilities between two independent servers: $\mathbb{S}_C$ dedicated to secure aggregation and $\mathbb{S}_E$ responsible for key generation, decryption, and model updates. In this architecture, the process begins with clients obtaining encryption keys, initial model parameters, and training hyperparameters (such as the optimizer and local learning rate) from $\mathbb{S}_E$ (Step $\circled{0}$ in \autoref{fig:fig_1}). Clients then encrypt their model updates and send the encrypted data to the aggregation server $\mathbb{S}_C$ (Step $\circled{1}$ in \autoref{fig:fig_1}).

After $\mathbb{S}_C$ aggregates these encrypted updates, the aggregated result is returned to $\mathbb{S}_E$ (Step $\circled{2}$ in \autoref{fig:fig_1}). $\mathbb{S}_E$ then decrypts the aggregated update, computes the updated model, and derives the reference update based on the aggregated result. Finally, $\mathbb{S}_E$ provides the decrypted model and the reference update back to the clients for further deployment or use (Step $\circled{3}$ in \autoref{fig:fig_1}). This process of Steps $\circled{1}$ to $\circled{3}$ is repeated until the pre-specified number of training epochs is completed.

 
\para{Advantage of \SAframework{}.}
Current SFL architectures typically adopt a single-server, such as masking-based~\cite{bonawitz2017practical, bell2020secure} and homomorphic encryption-based approaches~\cite{aono2017privacy, zhang2020batchcrypt, RoFL}. Masking-based methods are limited to addition aggregation, making them difficult to extend to Byzantine-robust aggregation techniques. Meanwhile, in homomorphic encryption-based schemes, if the server can impersonate a client, it could gain access to the encryption keys and decrypt collected local updates, exposing individual clients' private data. In real-world scenarios, preventing such server impersonation or ensuring that no client colludes with the server is a much stricter assumption than the one we make in \SAframework{}, which only requires that the two servers do not collude.

In \SAframework{}, secure aggregation operations are delegated to $\mathbb{S}_C$, while the remaining tasks are handled by $\mathbb{S}_E$. This ensures that $\mathbb{S}_C$ only sees encrypted intermediate results from clients without knowing the decryption key. Conversely, $\mathbb{S}_E$ can access the plaintext of the aggregated intermediate result, but this result is a sum of all clients' contributions, thus preventing the leakage of individual client-level data.


\para{Setup and Key Generation}.
In the \SAframework{} protocol, during the initial phase, the encryption server $\mathbb{S}_E$ generates an encryption key \texttt{ek} and a decryption key \texttt{dk}. It also initializes the model parameters $\beta_0$, selects a loss function $\ell$, an optimizer $\mathcal{O}$, a learning rate $\eta$, and the number of epochs $\mathbf{k}$. The server then sends \texttt{ek}, $\beta_0$, $\ell$, $\mathcal{O}$, and $\eta$ to each client, ensuring that the decryption key \texttt{dk} is not disclosed. This key is securely retained by $\mathbb{S}_E$ for use in decryption at a later stage.

\para{Client}.
Each client utilizes the homomorphic encryption public key $\texttt{ek}$ received from $\mathbb{S}_E$ to encrypt its local update $\mathbf{g}_i$. Let $n$ denote the length of the vector $\mathbf{g}_i$. The encryption of $\mathbf{g}_i$ is defined as follows: 
\[
\mathcal{C}(\mathbf{g}_i) = \left[ \mathcal{E}(\mathbf{g}_{i,1}, \texttt{ek}), \mathcal{E}(\mathbf{g}_{i,2}, \texttt{ek}), \dots, \mathcal{E}(\mathbf{g}_{i,n}, \texttt{ek}) \right]
\]

where each element of the vector is individually encrypted using $\texttt{ek}$. This ensures that $\mathbb{S}_C$ receives only the encrypted vector $\mathcal{C}(\mathbf{g}_i)$ without access to the original data $\mathbf{g}_i$, preserving the confidentiality of the client's information during transmission and storage.

\para{Computing Server $\mathbb{S}_C$}.
Once $\mathbb{S}_C$ receives the encrypted vectors $\mathcal{C}(\mathbf{g}_i)$, it can perform computations directly on the encrypted data, leveraging the homomorphic properties of the encryption scheme. The encrypted final model is then computed as:
\[
\mathcal{C}(\mathbf{g}) = \sum_{i=1}^m \mathcal{C}(\mathbf{g}_i)
\]
$\mathbb{S}_C$ subsequently transmits this encrypted output, $\mathcal{C}(\hat\beta)$, to the encryption server $\mathbb{S}_E$.

\para{Encryption Server $\mathbb{S}_E$}.
Upon receiving $\mathcal{C}(\hat\beta)$ from $\mathbb{S}_C$, $\mathbb{S}_E$ decrypts the aggregated vector to obtain the final computed model. The decryption process is expressed as:

\[
\mathbf{g} = \mathcal{D}(\mathcal{C}(\mathbf{g}), \texttt{dk})
\]

The new model parameters are updated as follows:
\[
\beta = \beta +  \alpha \cdot \frac{\mathbf{g} }{m}
\]

\noindent $\beta$ is then sent back to each client for new update iteration.

\section{ByzSFL: Byzantine-Robust Secure Federated Learning}
\label{sec:ByzSFL}

\begin{figure}[!t]
\centerline{\includegraphics[width=\linewidth]{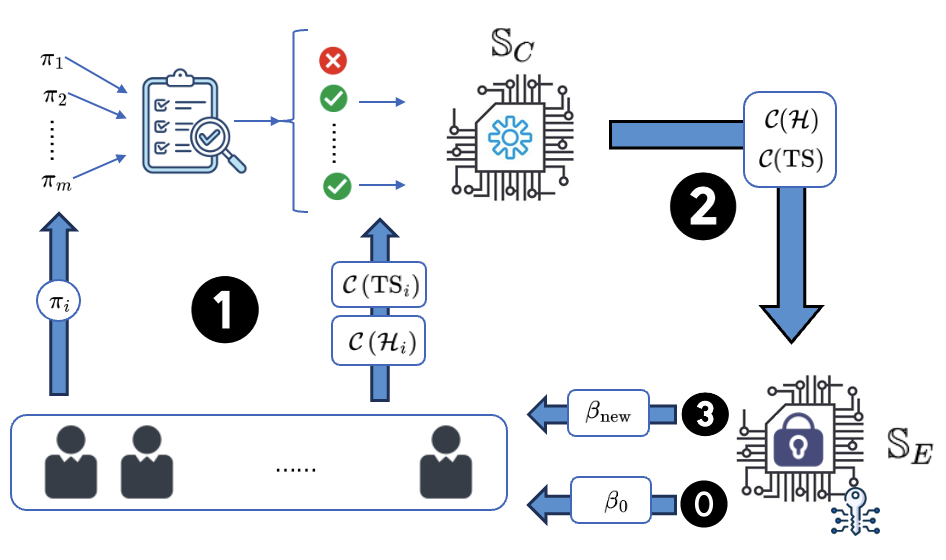}}
\caption{\small Framework of \ourmethod{}}
\label{fig:ByzSFL framework}
\end{figure}
\subsection{Design Goal}
The primary goal of \ourmethod{} is to adapt state-of-the-art Byzantine-robust input validation schemes from federated learning to secure federated learning. In this paper, we focus on FLTrust as our example due to its superior resilience in adversarial environments. Unlike other methods that typically fail when faced with high proportions of malicious clients, FLTrust can withstand over 90\% of such clients, significantly outperforming alternatives that generally handle only up to 50\%. This exceptional robustness makes it the ideal choice for our SFL framework. Moreover, FLTrust involves complex operations, such as cosine similarity, which require sophisticated Zero-Knowledge Proofs. Although our software also supports simpler validation schemes like Krum~\cite{Krum}, Trimmed Mean~\cite{Trim-mean}, and Median~\cite{Median}, we have omitted them from this discussion to keep the focus clear and avoid unnecessary complexity.

\subsection{Design Intuition}
The challenge in deploying FLTrust within a Secure Aggregation framework, as discussed earlier, lies in the fact that the input validation process involves complex operators. If FHE is used, it can handle these complex operations, but the overall system efficiency becomes significantly reduced due to the inherent computational overhead of FHE. On the other hand, while Partially PHE offers higher efficiency, it only supports one specific operations, such as addition (or multiplication), and cannot simultaneously handle both aggregation and input validation.

To address this limitation and enable the use of PHE while still supporting FLTrust’s input validation, we delegate the task of input validation to the clients themselves. Each client is responsible for validating its own update vector before uploading it. This approach allows clients to perform input validation locally, in plaintext, enabling quick calculations. Additionally, since each client performs its own validation independently, the process is parallelized across all clients, greatly enhancing efficiency.

However, allowing clients to validate their own update vectors introduces the risk that a client might accidentally or maliciously inflate its aggregation weight. To mitigate this risk, we have designed a ZKP protocol within \ourmethod{}. This protocol requires that, when a client submits its encrypted update vector and aggregation weight, it also generates and sends a ZKP. This proof demonstrates that the submitted aggregation weight has been correctly derived from the client’s update vector using the FLTrust validation process.

By integrating ZKP into the \ourmethod{} framework, we ensure that clients cannot falsely report their aggregation weights, thereby maintaining the integrity and robustness of the Secure Aggregation process while benefiting from the efficiency of PHE.

\begin{figure}[!t]
\begin{center}
\fbox{
\parbox{.95\linewidth}{
\textbf{Setup:}
\begin{enumerate}[itemsep=0pt]
    \item Server $\mathbb{S}_C$ sends initial model $\beta_0$, loss function $\ell$, optimizer $\mathcal{O}$, local learning rate $\eta$, global learning rate $\alpha$ and number of epoch $\mathbf{k}$ to each client.
    \item Server $\mathbb{S}_C$ compute $\beta^*$ by ~\autoref{eq:beta^*} and send to each client.
    \item Server $\mathbb{S}_E$ sends \texttt{ek} and \texttt{dk} to each client.
\end{enumerate}
}}\\
\fbox{
\parbox{.95\linewidth}{
\textbf{$i^{\text{th}}$ Client:}
\begin{enumerate}[itemsep=0pt]
    \item Compute $\beta_i$ by ~\autoref{eq:beta_i}
    \item Encrypt $\beta_i$ as $\mathcal{C}(\beta_i) = \mathcal{E}(\beta_i, \texttt{ek})$
    \item Send $\mathcal{C}(\beta_i)$ to server $\mathbb{S}_C$
\end{enumerate}
}}\\
\fbox{
\parbox{.95\linewidth}{
\textbf{Server $\mathbb{S}_C$}
\begin{enumerate}[itemsep=0pt]
    \item Compute $\mathcal{C}(\mathbf{g}) =  \sum_{i=1}^m \mathcal{C}(\mathbf{g}_i)$
    \item Send $\mathcal{C}(\mathbf{g})$ to server $\mathbb{S}_E$
\end{enumerate}
}}\\ 
\fbox{
\parbox{.95\linewidth}{
\textbf{Server $\mathbb{S}_E$}
\begin{enumerate}[itemsep=0pt]
    \item Decrypt to get $\mathbf{g} = \mathcal{D}(\mathcal{C}(\mathbf{g}), \texttt{dk})$
    \item $\beta = \beta +  \alpha \cdot \frac{\mathbf{g} }{m}$
    \item Send $\beta$ to each client
\end{enumerate}
}}
\caption{Protocol details of \SAframework{}}
\label{fig:protocol}
\end{center}
\end{figure}

\subsection{Design Overview}

In the \ourmethod{} protocol, each client $i$ computes their own $\text{TS}_i$, $\Tilde{\text{TS}}_i$, and $\mathcal{H}_i$ and encrypted as $\mathcal{C}(\mathcal{H}_i)$ and $\mathcal{C}(\text{TS}_i)$. Then, each client generates a zk-SNARK proof to demonstrate the integrity and correctness of their computations. The workflow of the protocol framework is presented in \autoref{fig:ByzSFL framework}, showing distinct steps and transmissions compared to the general framework of~\SAframework{}, which is depicted in \autoref{fig:fig_1}. Additionally, we provide the detailed protocol design in \autoref{fig:phe_protocol}. Note that, given that $\mathbb{S}_C$ in the \SAframework{} architecture can only perform addition, we have optimized the distribution of computations for the FLTrust component section shown in \autoref{fig:phe_protocol}). Our adapted distribution maintains mathematical equivalence with original formulation of FLTrust, ensuring efficient and accurate outcomes.

Subsequently, Server $\mathbb{S}_C$ only needs to perform partial homomorphic encryption with an addition operation. Note that the number of clients being proved is a constant number, which does not require multiplication operations in homomorphic encryption.


For the encryption, We selected the Paillier cryptosystem, a partial homomorphic encryption scheme that supports homomorphic addition over encrypted integers, allowing for computations on ciphertexts without decrypting the data. We also integrate zk-SNARKs to ensure the integrity and correctness of the client's computations. The detailed ZKP protocol will be discussed in \autoref{sec:zkp}.

\subsection{Zero Knowledge Proof}
\label{sec:zkp}
\para{ZKP Protocol Selection}.
Zero-knowledge proofs have been employed in various machine learning and artificial intelligence research projects to enhance data privacy by enabling computations on encrypted data without revealing the underlying information~\cite{zhu2023robust, bell2023acorn, han2023kick}. For example, Bulletproofs~\cite{bunz2018bulletproofs}, a non-interactive zero-knowledge proof system, have been used to prove whether a vector norm exceeds a pre-set threshold. However, Bulletproofs have certain limitations. While they are more efficient and easier to implement than some other zero-knowledge proof systems for inequality statements, especially regarding proof size, they demand more computational resources for verification. This requirement can make them less suitable for applications that require fast verification times or operate on resource-constrained devices. Furthermore, the complexity of Bulletproofs can increase significantly with the size of the computation, as their proof size scales with $O(\log n)$ and their verification time with $O(n)$.

In contrast, zk-SNARKs offer significant advantages over Bulletproofs, particularly in terms of communication complexity (proof size) and verification speed. The verification time and proof size of zk-SNARKs are independent of the size of the statement, consistently requiring $O(1)$ for both. When dealing with complex and large statements, zk-SNARKs reduce communication costs and offer faster verification.

Several zk-SNARK protocols have been developed to optimize proof generation and verification efficiency, with notable examples including Pinocchio~\cite{parno2016pinocchio} and Groth16~\cite{groth2016size}. Pinocchio was one of the first practical zk-SNARK protocols, providing efficient proof generation and verification while maintaining a relatively small proof size. However, it still required 12 elliptic curve pairing operations during verification, which could become computationally intensive as the complexity of the computation increased. Groth16, on the other hand, represents a significant advancement in zk-SNARK protocols. It further reduces both the proof size and the complexity of the verification process. Specifically, Groth16 generates a proof that consists of only three elliptic curve group elements. This compact proof structure allows for faster verification, requiring just four elliptic curve pairing operations regardless of the complexity of the underlying computation. These features make Groth16 one of the most efficient zk-SNARK protocols available, particularly suitable for applications where minimizing proof size and verification time is crucial.

Many dataset such as MNIST contain a large number of parameters, resulting in a large R1CS circuit and input. Therefore, we selected zk-SNARKs with the Groth16 protocol to minimize proof size and verification time.

\begin{figure}[!t]
\begin{center}
\fbox{
\parbox{.95\linewidth}{
\textbf{Setup:}
\begin{enumerate}[itemsep=0pt]
    \item Server $\mathbb{S}_E$ sends initial model $\beta_0$, validation dataset $D^*$, loss function $\ell$, optimizer $\mathcal{O}$, local learning rate $\eta$, global learning rate $\alpha$ and number of epoch $\mathbf{k}$ to each client.
    \item Server $\mathbb{S}_E$ compute $\mathbf{g}^*$ by ~\autoref{eq:beta^*} and send to each client.
    \item Server $\mathbb{S}_E$ generate homomorphic encryption keys \texttt{ek}, \texttt{dk} and ZKP keys \texttt{pk}, \texttt{vk}
    \item Server $\mathbb{S}_E$ sends \texttt{ek}, \texttt{pk} to each client and \texttt{vk} to $\mathbb{S}_C$ 

\end{enumerate}
}}
\fbox{
\parbox{.95\linewidth}{
\textbf{$i^{\text{th}}$ Client:}
\begin{enumerate}[itemsep=0pt]
    \item Compute $\mathbf{g}_i$ by ~\autoref{eq:beta_i}
    \item Compute $\text{TS}_i$ and  $\Tilde{\text{TS}}_i$ by ~\autoref{eq:fltrust score} and ~\autoref{eq:normalizd fltrust score}
    \item Compute $\mathcal{H}_i = \Tilde{\text{TS}}_i \cdot \mathbf{g}_i$
    \item Encrypt $\mathcal{C}(\mathcal{H}_i) = \mathcal{E}(\mathcal{H}_i, \texttt{ek})$
    \item Encrypt $\mathcal{C}(\text{TS}_i) = \mathcal{E}(\text{TS}_i, \texttt{ek})$ 
    \item Produce proof $\pi_i = \texttt{Prove}(\mathbf{g}_i, \mathcal{C}(\mathcal{H}_i), \mathcal{C}(\text{TS}_i), \texttt{pk})$
    \item Send $\pi_i$, $\mathcal{C}(\text{TS}_i)$ and $\mathcal{C}(\mathcal{H}_i)$ to server $\mathbb{S}_C$
\end{enumerate}
}}
\fbox{
\parbox{.95\linewidth}{
\textbf{Server $\mathbb{S}_C$:}
\begin{enumerate}[itemsep=0pt]
    \item Verify all proofs $\pi_i$ received from each client with $\texttt{Prove}(\pi_i, \texttt{vk})$; reject $i^{th}$ if verification is fail
    \item Compute $\mathcal{C}(\mathcal{H}) = \sum_{i \in \text{Proved clients}} \mathcal{C}(\mathcal{H}_i)$
    \item Compute $\mathcal{C}(\text{TS}) = \sum_{j \in \text{Proved clients}}\mathcal{C}(\text{TS}_j)$
    \item Send $\mathcal{C}(\mathcal{H})$ and $\mathcal{C}(\text{TS})$ to server $\mathbb{S}_E$
\end{enumerate}
}}
\fbox{
\parbox{.95\linewidth}{
\textbf{Server $\mathbb{S}_E$:}
\begin{enumerate}[itemsep=0pt]
    \item Decrypt $\mathcal{C}(\mathcal{H})$ and $\mathcal{C}(\text{TS})$
    \item Compute new global model $\beta_{\text{new}} = \beta_{\text{new}} + \alpha\cdot  \frac{\mathcal{H}}{\text{TS}}$
    \item Send the updated global model $\beta_{\text{new}}$ to each client for next round until completing $\mathbf{k}$ epochs.
\end{enumerate}
}}

\caption{Protocal of \ourmethod{}}
\label{fig:phe_protocol}
\end{center}
\end{figure}

\para{zk-SNARKs for Federated Learning}.
Build a zk-SNARK circuit and proof for secure aggregation presents significant challenges due to the inherent complexity of both zk-SNARKs and secure federated learning models. Deep learning models are typically large and complex, with numerous parameters and operations that need to be encoded into zk-SNARK-compatible arithmetic circuits. Federated learning models usually are difficult to represent in a zk-SNARK-friendly way such as use floating-point arithmetic, while zk-SNARKs require integer or fixed-point arithmetic, necessitating complex conversions that can affect model precision. Due to the complexity of encoding machine learning models in this way, zk-SNARKs for federated learning models generate a high computational overhead, leading to lengthy proof generation times that may not be practical for frequent updates typical in federated learning scenarios. 

Currenly designed zero knowledge proof for federated learning in previous research were only produce a proof for some small and simply statement such as range and bound proving. Therefore, it is hard to apply the previous development into our model. There are tools like gadgets and languages like Circom that assist in building zk-SNARK circuits by simplifying and modularizing complex operations. Gadgets are reusable components designed for common cryptographic functions, such as hash computations or basic arithmetic, making it easier to construct circuits by piecing together these building blocks. However, while these tools are useful for many zero-knowledge applications, they currently lack the flexibility and scale needed to support federated learning models. 

To simplify the construction of zero-knowledge proofs for federated learning models, we developed a new gadget library based on Circom. This library modularizes complex operations frequently used in Byzantine-robust technique, making circuit construction more manageable. The library includes the following gadgets, which can be used individually or in combination:
\begin{enumerate}[itemsep=0pt]
    \item Vector computation with floating-point numbers
    \item Vector median and mean
    \item $L_{2}$ norm computation
    \item Cosine similarity computation
    \item Range comparison and min/max functionality
    \item Additive homomorphic encryption
    \item Multiplicative homomorphic encryption
\end{enumerate}
With these gadgets, developers can efficiently construct zk-SNARK circuits tailored for federated learning models. In Section \ref{subsec:zk-SNARKs Construction}, we demonstrate how to use our newly developed gadget library to generate a zk-SNARK circuit for our secure federated learning model. To construct these gadgets, we encode these mathematical operations into the Circom ~\cite{belles2022circom} language, utilizing the Groth16 protocol and BLS12-381 curve, a pairing-friendly elliptic curve defined by the equation $y^2 = x^3 + 4$ and $r = \seqsplit{52435875175126190479447740508185965837690552500527637822603658699938581184513}$. Circom allows us to define custom circuits that represent the necessary arithmetic and logical operations for our proof, and it assists in converting the proving statement into an R1CS circuit. Since the zk-SNARK protocol operates over finite fields rather than real numbers, it does not support floating-point arithmetic. Therefore, we apply fixed-point arithmetic to convert all data into integers instead of floating-point numbers.

\para{zk-SNARKs Construction}.
\label{subsec:zk-SNARKs Construction}
Now we are discussing how to construction the zk-SNARKs circuit in our protocol. Firstly, we need to design and implement the zk-SNARK circuit. To achieve the goal of allowing the client to self-prove the integrity and correctness of computations using zero knowledge proof, we have divided the zk-SNARKs design into three stages. The client’s submission will be considered valid if and only if it successfully passes all three stages. In simple terms, these three stages are: a proof of the integrity of $\text{TS}_i$ and $\tilde{\text{TS}}_i$, a proof of the correctness of the computation of $\mathcal{H}_i$, and a proof of the accuracy of the encryption process for $\text{TS}_i$ and $\mathcal{H}_i$. Specifically, the client must first 
uses cosine similarity and $L_{2}$ norm computation gadgets to prove that $\text{TS}_i$ and $\tilde{\text{TS}}_i$ are computed as
\[
\text{TS}_i = \max(0, \frac{\bold{g}^* \cdot \bold{g}_i}{\|\bold{g}^*\| \|\bold{g}_i\|})
\]
\[
\tilde{\text{TS}}_i =  \text{TS}_i \cdot \frac{\left\|\bold{g}^*\right\|}{\left\|\bold{g}_i\right\|}
\]
Next, the client needs to use the vector computation gadget to demonstrate that
\[
\mathcal{H}_i = \text{TS}_i \cdot \mathbf{g}_i = \left[\text{TS}_i \cdot \mathbf{g}_{i,1}, \text{TS}_i \cdot \mathbf{g}_{i,2}, \dots, \text{TS}_i \cdot \mathbf{g}_{i,n} \right]
\]
Finally, since the client will not send the unencrypted data to the server, they need to use additive homomorphic encryption gadget to prove that
\[
\mathcal{C}(\mathcal{H}_i) = \mathcal{E}(\mathcal{H}_i, \texttt{ek})
\]
\[
\mathcal{C}(\text{TS}_i) = \mathcal{E}(\text{TS}_i, \texttt{ek})
\]

After construct the circuit $C$ with gadgets mentioned above, the client then submits a proof that these constraints hold true without revealing the private inputs. Witness $\vec{a}$ is a set of private inputs and intermediate values that satisfy the constraints of the circuit, allowing the prover to generate a proof that the computation was carried out correctly without revealing the actual data. In our zk-SNARKs circuit, $\mathbf{g}_i$, $\text{TS}_i$, $\Tilde{\text{TS}}_i$, and $\mathcal{H}_i$ are private inputs. The public input, on the other hand, consists of the values that are made available to both the prover and the verifier. These inputs are included in the proof and are necessary for the verification process.

After implementing the necessary operations in Circom, the next step is to construct the zk-SNARK proof system around the circuit. This involves compiling the Circom code into a constraint system that can be utilized by the zk-SNARK protocol. The compiled circuit $C$ is then used to generate a proving key and a verification key, both of which are essential components of the zk-SNARK system. The proving key enables the prover to generate a proof that the circuit's constraints are satisfied for the given inputs. Conversely, the verification key allows the verifier to check the validity of the proof without gaining any knowledge of the private inputs. This process ensures that the verifier can confidently trust the correctness of the computation without needing access to the actual data. The detailed procedure is shown in \autoref{fig:zkp_protocol}.

zk-SNARKs is a non-interactive zero knowledge proof with trust setting up require. The setup phase generates public parameters that are essential for both proof generation and verification. During this phase, the trusted setup also produces the proving key (\texttt{pk}) and the verification key (\texttt{vk}). Both the proving key and the verification key are public key and can be disclosed to any party. In our protocol, we designate Server $\mathbb{S}_E$ to perform the trusted setup, as it is also responsible for generating the homomorphic encryption and decryption keys. Server $\mathbb{S}_E$ does not receive the original encrypted or unencrypted client data and does not act as either the prover or the verifier in the protocol. The setup phase is crucial for securely establishing the cryptographic structure required for zero-knowledge proofs. It is important to note that for the same statement that a prover wants to prove with different input values, zk-SNARKs require only one universal setup, allowing all parameters and keys to be reused multiple times by different parties without new setup. 

\texttt{snarkjs} is a JavaScript library designed for working with zk-SNARKs, providing the \texttt{Setup}, \texttt{Prove}, and \texttt{Verify} functions. It enables proof generation by the client as the prover and proof verification by the server as the verifier. If the proof is valid, it confirms that the computation was carried out correctly according to the defined circuit, without revealing any of the secret inputs or intermediate results.


\subsection{Privacy and Security Analysis}
\label{subsec:Privacy and Security Analysis}

In our framework, secure aggregation operations are delegated to $\mathbb{S}_C$, while the remaining tasks are handled by $\mathbb{S}_E$. This design ensures that $\mathbb{S}_C$ only processes encrypted intermediate results from clients without having access to the decryption key. Meanwhile, $\mathbb{S}_E$ can view the plaintext of the aggregated results, but since these results are a sum of all clients' contributions, individual client-level data remains protected.

\textbf{Collusion Among Clients:} In our framework, collusion between clients does not pose a threat to privacy. Each client only has access to its own encrypted intermediate results and has no information about other clients' data. Consequently, even if clients attempt to collaborate, they would not gain any additional insight into the private data of others.

\textbf{Collusion Between Clients and $\mathbb{S}_C$ or $\mathbb{S}_E$:} Similarly, privacy is not compromised if any client colludes with either $\mathbb{S}_C$ or $\mathbb{S}_E$. $\mathbb{S}_C$ only handles encrypted data and does not have the decryption key, while $\mathbb{S}_E$ only sees the aggregated result, which is protected by the secure aggregation protocol. Therefore, even in the event of such collusion, individual contributions from other clients remain secure. Meanwhile, validation is not compromised if $\mathbb{S}_C$ does not exclude clients whose zero-knowledge proofs cannot be verified.

\textbf{Collusion Between $\mathbb{S}_C$ and $\mathbb{S}_E$:} Firstly, if  $\mathbb{S}_C$  colludes with a client, it may ignore invalid ZKP proofs submitted by the colluding client, treating them as valid, thereby introducing a vulnerability in the system. Furthermore, when  $\mathbb{S}_C$  and  $\mathbb{S}_E$  collude $\mathbb{S}_C$ could share the encrypted intermediate results with $\mathbb{S}_E$, who, with the decryption key, could decrypt these results and reveal individual clients' local updates. However, we contend that collusion between $\mathbb{S}_C$ and $\mathbb{S}_E$ is more manageable to prevent and monitor in practice. This is because $\mathbb{S}_C$ and $\mathbb{S}_E$ are generally operated by independent and mutually distrustful entities, such as different organizations or divisions within a company. Additionally, implementing regulatory measures, contractual agreements, and auditing mechanisms can effectively deter and detect any unauthorized collaboration between these servers, significantly lowering the risk of such collusion.

\textbf{Clients Sending Arbitrary Model Updates:} In our framework, clients are required to follow the protocol for computing aggregation weights, as any deviation would be detected and filtered out by the ZKP procedure. Nonetheless, clients could still submit arbitrary local model updates. If a client chooses to do this, the resulting cosine similarity between their update and the reference update would be very low. Consequently, the client would receive a significantly reduced aggregation weight, greatly minimizing their impact on the overall model update. This mechanism ensures that even if a client behaves maliciously by sending arbitrary updates, their influence on the aggregated model is negligible, thereby preserving the integrity and robustness of the training process.

\begin{figure}[!t]
\centering
\fbox{
\parbox{.95\linewidth}{
\textbf{Public parameter:} A prime $r$, two cyclic groups $\mathbb{G}_1$ and $\mathbb{G}_2$ of order $r$ with generator $g_1 \in \mathbb{G}_1$ and $g_2 \in \mathbb{G}_2$, a pairing $e: \mathbb{G}_1 \times \mathbb{G}_2 \rightarrow \mathbb{G}_T$ of order $r$.
}}
\fbox{
\parbox{.95\linewidth}{
\textbf{Server $\mathbb{S}_E$ (setup):}\\
INPUT: circuit $C \in \mathbb{F}_r^{l}$\\
OUTPUT: proving key \texttt{pk}, verification key \texttt{vk}
\begin{enumerate}[itemsep=0pt]
    \item Generate circuit $C$ with \texttt{Circom}
    \item Generate key $\texttt{pk},\texttt{vk} = \texttt{Setup}(C)$
\end{enumerate}
}}
\fbox{
\parbox{.95\linewidth}{
\textbf{Client $i$ (prover):}\\
INPUT: proving key \texttt{pk}, witness $\vec{a}$, public input $\vec{x}$\\
OUTPUT: proof $\pi$
\begin{enumerate}[itemsep=0pt]
    \item Set witness $\vec{a} = \mathbf{g}_i \Vert \text{TS}_i \Vert \Tilde{\text{TS}}_i \Vert \mathcal{H}_i$
    \item Set public input $\vec{a} = \bold{g}^*$
    \item Generate proof $\pi = \texttt{Prove}(\texttt{pk}, \vec{a}, \vec{x})$ \\ $\pi$ contains two points $\in \mathbb{G}_1$ and one point $\in \mathbb{G}_2$
\end{enumerate}

}}
\fbox{
\parbox{.95\linewidth}{
\textbf{Server $\mathbb{S}_C$ (verifier):}\\
INPUT: verification key \texttt{vk}, proof $\pi$, public input $\vec{x}$\\
OUTPUT: verification result in true/false
\begin{enumerate}[itemsep=0pt]
    \item Verify the proof $\texttt{Bool} = \texttt{Verify}(\texttt{vk}, \pi, \vec{x})$
    \item Accept if verification result is true
\end{enumerate}
}}
\caption{Detailed protocol of zk-SNARKs for each party (client, server $\mathbb{S}_C$, and server $\mathbb{S}_E$)}
\label{fig:zkp_protocol}
\end{figure}

\section{Implementation and Evaluation}
\label{sec:Implementation and Evaluation}


In this section, we evaluate our approach using the MNIST dataset across various benchmarks to assess its accuracy and efficiency. We compare our method with several existing techniques, highlighting its relative performance in terms of time complexity and transmission cost, thereby demonstrating the strengths of our approach.

\begin{table*}[!t]
\setlength{\tabcolsep}{7mm}{
\begin{center}
\small
\begin{tabular}{c||cccccc}
\toprule
 \multirow{2}{*}{Step} & \multicolumn{6}{c}{Number of Parameters} \\ \cmidrule{2-7} 
& \multicolumn{1}{c|}{9k} & \multicolumn{1}{c|}{19k} & \multicolumn{1}{c|}{38k} & \multicolumn{1}{c|}{76k} & \multicolumn{1}{c|}{152k} & 304k  \\ 
\midrule
Client Compute        & \multicolumn{1}{c|}{0.07}  & \multicolumn{1}{c|}{0.07} & \multicolumn{1}{c|}{0.08}  & \multicolumn{1}{c|}{0.08}& \multicolumn{1}{c|}{0.09}      & 0.11 \\ 
Client Encrypt        & \multicolumn{1}{c|}{3.25}  & \multicolumn{1}{c|}{5.56} & \multicolumn{1}{c|}{10.9}  & \multicolumn{1}{c|}{20.17}& \multicolumn{1}{c|}{41.80}      & 75.99 \\ 
Client Prove          & \multicolumn{1}{c|}{0.99}  & \multicolumn{1}{c|}{1.27} & \multicolumn{1}{c|}{2.03}  & \multicolumn{1}{c|}{3.42}& \multicolumn{1}{c|}{5.04}      & 8.43 \\ \midrule
Server $\mathbb{S}_C$ Compute      & \multicolumn{1}{c|}{0.09}  & \multicolumn{1}{c|}{0.13} & \multicolumn{1}{c|}{0.24}  & \multicolumn{1}{c|}{0.46}& \multicolumn{1}{c|}{0.92}      & 2.01 \\ 
Server $\mathbb{S}_C$ Verify       & \multicolumn{1}{c|}{0.79}  & \multicolumn{1}{c|}{0.75} & \multicolumn{1}{c|}{0.74}  & \multicolumn{1}{c|}{0.74}& \multicolumn{1}{c|}{0.76}      & 0.73 \\
Server $\mathbb{S}_E$ Decrypt      & \multicolumn{1}{c|}{3.11}  & \multicolumn{1}{c|}{5.13} & \multicolumn{1}{c|}{11.25}  & \multicolumn{1}{c|}{20.34}& \multicolumn{1}{c|}{40.12}      & 77.83 \\ \midrule
Total      & \multicolumn{1}{c|}{8.51}  & \multicolumn{1}{c|}{13.08} & \multicolumn{1}{c|}{15.89}  & \multicolumn{1}{c|}{46.27}& \multicolumn{1}{c|}{89.52}      & 166.41 \\ 
\bottomrule
\end{tabular}
\end{center}
\caption{Time performance evaluation results based on the different number of parameter in dataset}
\label{tab:vector_length}
}
\end{table*}

\begin{table*}[!t]
\setlength{\tabcolsep}{6mm}{
\begin{center}
\small
\begin{tabular}{c||ccccccc}
\toprule
\multirow{2}{*}{Step} & \multicolumn{7}{c}{Client Size} \\ \cmidrule{2-8} 
& \multicolumn{1}{c|}{2}  & \multicolumn{1}{c|}{4}  & \multicolumn{1}{c|}{8}  & \multicolumn{1}{c|}{16} & \multicolumn{1}{c|}{32} & \multicolumn{1}{c|}{64} & \multicolumn{1}{c}{128} \\ \midrule
Server $\mathbb{S}_C$ Compute      & \multicolumn{1}{c|}{0.13} & \multicolumn{1}{c|}{0.41} & \multicolumn{1}{c|}{0.99} & \multicolumn{1}{c|}{2.13} & \multicolumn{1}{c|}{5.27} & \multicolumn{1}{c|}{9.88} & 18.07                       \\  
Server $\mathbb{S}_C$ Verify       & \multicolumn{1}{c|}{0.78} & \multicolumn{1}{c|}{1.51} & \multicolumn{1}{c|}{2.96} & \multicolumn{1}{c|}{5.89} & \multicolumn{1}{c|}{11.84} & \multicolumn{1}{c|}{23.69} & 47.29                       \\  
Server $\mathbb{S}_E$ Decrypt      & \multicolumn{1}{c|}{4.97} & \multicolumn{1}{c|}{5.19} & \multicolumn{1}{c|}{4.81} & \multicolumn{1}{c|}{5.02} & \multicolumn{1}{c|}{5.17} & \multicolumn{1}{c|}{5.15} & 5.10  \\ 
\bottomrule
\end{tabular}
\end{center}
\caption{Time performance evaluation results based on the different number of clients}
\label{tab:client_number}
}
\end{table*}

\subsection{Experimental Setup}
To evaluate the performance of our protocol with partially homomorphic encryption (PHE), we also developed a protocol using fully homomorphic encryption (FHE), which is introduced in \autoref{subsec:baseline}. We conducted several experiments to demonstrate improved performance in both time and space complexity. 

We implemented both PHE and FHE schemes in Python 3, constructed the zk-SNARKs circuit using \texttt{Circom}, and performed zk-SNARK proving and verification with \texttt{snarkJS}. All evaluation experiments were conducted on a machine with an Intel Core i7-8700 CPU @ 3.20GHz × 12, 32GB of RAM, and running 64-bit Ubuntu 22.04 LTS.

\para{Baseline Protocol with FHE}.
\label{subsec:baseline}
We also developed a baseline protocol using fully homomorphic encryption (FHE). We selected the Cheon, Kim, Kim, and Song (CKKS) scheme, a fourth-generation FHE method that enables efficient computation on encrypted data for both addition and multiplication while supporting operations on floating-point numbers with controlled precision loss. The detailed protocol is presented in Appendix~\ref{sec:fhe_protocol}.

\subsection{Efficacy Evaluation of \ourmethod}
\label{subsec:Efficacy}
\para{Impact of Number of Parameters}.
First, we evaluate the end-to-end performance of our protocol with varying numbers of parameters. In this setup, the server receives data from two clients, focusing on the effect of parameter variation. The impact of increasing the number of clients will be assessed in a subsequent evaluation. \autoref{tab:vector_length} presents the evaluation results for each step in our protocol.
Since clients operate in parallel—computing, encrypting, and generating zk-SNARK proofs simultaneously—we report the time taken by the slowest client. All runtimes in \autoref{tab:vector_length} are measured in seconds.


The primary component of the runtime is consumed by the processes of vector encryption and decryption. The time required for both encryption and decryption grows linearly with the number of parameters, which corresponds to the increasing size of the vector. Consequently, as the vector size expands, the encryption and decryption workload increases proportionally. Cryptographic algorithms process data in fixed-size blocks or with consistent operations, ensuring that computational effort remains unaffected by the actual values within the data. Thus, the encryption time depends solely on the vector size, not the data values.

The zk-SNARK proof verification time on Server $\mathbb{S}_C$ remains constant as the number of parameters increases because zk-SNARK proof verification has a time complexity of $O(1)$, independent of input size and statement complexity. This ensures high efficiency for Server $\mathbb{S}_E$ when verifying the correctness of clients' computations. While our scheme requires additional time for clients to generate zk-SNARK proofs and for Server $\mathbb{S}_C$ to verify them, partial homomorphic encryption is approximately 85 times faster than fully homomorphic encryption schemes. Consequently, employing partial homomorphic encryption alongside zero-knowledge proofs offers significantly better performance compared to using fully homomorphic encryption alone. In \autoref{subsec:comparison}, we provide a comparison between our proposed scheme and a fully homomorphic encryption scheme, highlighting the computational efficiency and performance advantages of our approach.

\para{Impact of Number of Clients}.
As the number of clients increases, Server $\mathbb{S}_C$ must perform more computations on encrypted data and verify a larger number of zk-SNARK proofs, which can impact its overall runtime. These additional computational demands result in longer processing times. To better understand our protocol's performance with varying numbers of clients, we conducted a comprehensive evaluation to assess how the server’s processing time changes as the number of clients increases. We set the number of parameters to 19K, and \autoref{tab:client_number} presents the evaluation results for 2, 4, 8, 16, 32, 64, and 128 clients. All runtimes in \autoref{tab:client_number} are measured in seconds.


As the number of clients increases, Server $\mathbb{S}_C$ must process a growing volume of computations on encrypted vectors under homomorphic encryption. Each additional client contributes an encrypted input requiring secure processing, leading to a linear increase in computational demand. The zk-SNARK proof verification time for Server $\mathbb{S}_C$ also increases with the number of clients, as more proofs need to be verified.

Server $\mathbb{S}_E$ receives only the encrypted aggregated update, meaning the size of the update does not increase with the number of clients. As a result, the decryption time remains unaffected by the number of clients. Additionally, $\mathbb{S}_E$  computes the new global model, but this computation is performed on a plain vector rather than an encrypted one, resulting in negligible processing time.

\para{Communication Cost}.
Since our scheme involves interaction between the client, Server $\mathbb{S}_C$, and Server $\mathbb{S}_E$, the size of data transmission is a critical factor, as large transmissions can increase latency. To evaluate this aspect, we conducted an experiment to measure the transmission costs within our scheme. Specifically, we measured the data sent from each client to Server $\mathbb{S}_C$, from Server $\mathbb{S}_C$ to Server $\mathbb{S}_E$, and from Server $\mathbb{S}_E$ back to each client. Notably, the number of clients does not affect the size of the vector transmitted from Server $\mathbb{S}_C$ to Server $\mathbb{S}_E$. Therefore, our evaluation focuses on how transmission costs vary with different vector sizes. \autoref{tab:transmission} presents the transmission cost for each party, including the size of zk-SNARK proofs and the vector. In \autoref{subsec:comparison}, we will also compare the transmission costs of our scheme utilizing partially homomorphic encryption with those of a protocol using fully homomorphic encryption.

\begin{table*}[!t]
\setlength{\tabcolsep}{1.8mm}{
\begin{center}
\small
\begin{tabular}{cc||cccccc}
\toprule
\multicolumn{2}{c||}{\multirow{2}{*}{Transmission}} & \multicolumn{6}{c}{Number of Parameters} \\ \cmidrule{3-8} 
& & \multicolumn{1}{c|}{9k} & \multicolumn{1}{c|}{19k} & \multicolumn{1}{c|}{38k} & \multicolumn{1}{c|}{76k} & \multicolumn{1}{c|}{152k} & 304k \\ 
\midrule
\multirow{2}{*}{Client to Server $\mathbb{S}_C$} & \multicolumn{1}{|c||}{Encrypted Vector} & \multicolumn{1}{c|}{4.42 MB}  & \multicolumn{1}{c|}{9.08 MB} & \multicolumn{1}{c|}{17.72 MB}  & \multicolumn{1}{c|}{36.52 MB}& \multicolumn{1}{c|}{64.86 MB}      & 127.83 MB \\ 
& \multicolumn{1}{|c||}{ZKP Proof} & \multicolumn{1}{c|}{803 Bytes}  & \multicolumn{1}{c|}{803 Bytes} & \multicolumn{1}{c|}{803 Bytes}  & \multicolumn{1}{c|}{803 Bytes}& \multicolumn{1}{c|}{803 Bytes}      & 803 Bytes \\ \midrule
Server $\mathbb{S}_C$ to Server $\mathbb{S}_E$ & \multicolumn{1}{|c||}{Encrypted Vector} & \multicolumn{1}{c|}{4.42 MB}  & \multicolumn{1}{c|}{9.08 MB} & \multicolumn{1}{c|}{17.72 MB}  & \multicolumn{1}{c|}{36.52 MB}& \multicolumn{1}{c|}{64.86 MB}      & 127.83 MB \\ \midrule
Server $\mathbb{S}_E$ to Client & \multicolumn{1}{|c||}{Vector} & \multicolumn{1}{c|}{0.31 MB}  & \multicolumn{1}{c|}{0.62 MB} & \multicolumn{1}{c|}{1.22 MB}  & \multicolumn{1}{c|}{2.45 MB}& \multicolumn{1}{c|}{4.89 MB}      & 9.71 MB \\ 
\bottomrule
\end{tabular}
\end{center}
\caption{Transmission cost for each communication between client, server $\mathbb{S}_C$, and server $\mathbb{S}_E$}
\label{tab:transmission}
}
\vspace{10pt}
\end{table*}

\subsection{Comparison of \ourmethod{} with Other Schemes}
\label{subsec:comparison}
\begin{figure}[!t]
\centerline{\includegraphics[width=\linewidth]{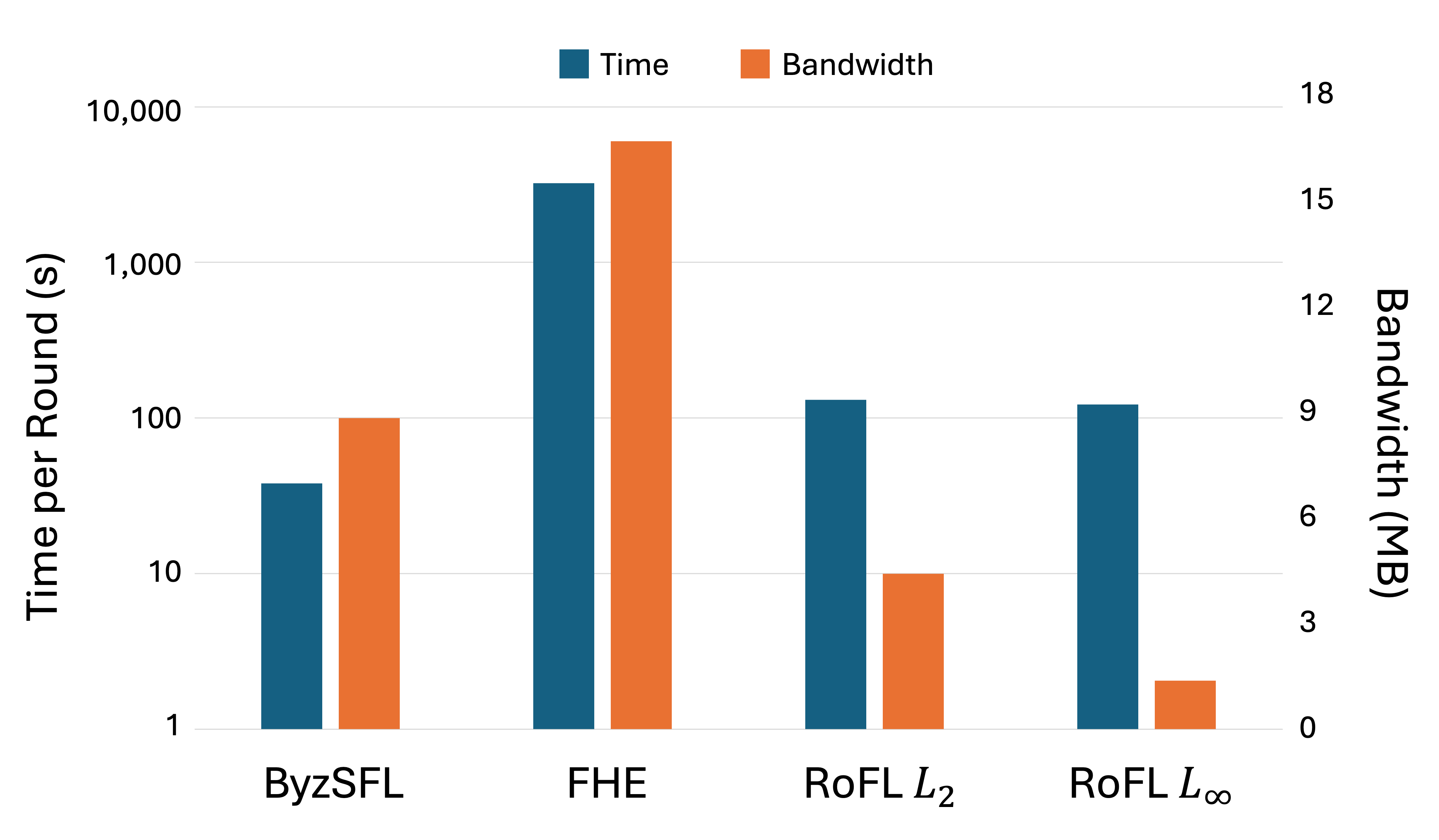}}
\caption{Runtime and Bandwidth Comparison Between \ourmethod{} and other schemes}
\label{fig:comparison}
\end{figure}

We compare the performance and key characteristics of our proposed method against two benchmarks: the baseline protocol that employs fully homomorphic encryption and the RoFL approach with \( L_{2} \) and \( L_{\infty} \) norm regularization. Figure \ref{fig:comparison} shows a comparison of time performance in seconds (left y-axis) and bandwidth usage (right y-axis).

The comparison uses the same metrics, with 19K parameters and the MNIST dataset. We evaluate the differences in time performance and transmission cost between each scheme.

We present a fully homomorphic encryption scheme in Appendix~\ref{sec:fhe_protocol} as a baseline that shares structural similarities with \ourmethod{}. Since fully homomorphic encryption enables both addition and multiplication, the client is not required to perform computations defined by \autoref{eq:fltrust score} and \autoref{eq:normalizd fltrust score}. These computations are instead performed by $\mathbb{S}_C$, leading to additional computation time for $\mathbb{S}_C$, particularly as the number of clients increases. Consequently, since the client does not perform these computations, there is no need for the client to generate zero-knowledge proofs to verify their computations.

Based on our evaluation, as shown in \autoref{fig:comparison}, \ourmethod{}, which utilizes zk-SNARKs for faster proving and verification along with partially homomorphic encryption, achieves approximately 85 times the speed of the baseline protocol employing fully homomorphic encryption. Specifically, our scheme requires only 38 seconds per round, while the fully homomorphic encryption protocol requires 3224 seconds per round. Additionally, RoFL with \( L_{2} \) norm regularization requires 131 seconds, and RoFL with \( L_{\infty} \) norm regularization requires 122 seconds.

The fully homomorphic encryption scheme demonstrates poorer performance compared to \ourmethod{} for several reasons. First, fully homomorphic encryption requires significantly longer encryption and decryption runtimes compared to partially homomorphic encryption. Second, unlike \ourmethod{}, where clients compute their local updates in parallel, $\mathbb{S}_C$ must handle encrypted computations for each client individually. Lastly, the cost of generating zero-knowledge proofs in \ourmethod{} is considerably lower than the additional runtime incurred by fully homomorphic encryption. These advantages make partially homomorphic encryption combined with zero-knowledge proofs a more efficient solution than fully homomorphic encryption.

Furthermore, \ourmethod{} outperforms both RoFL with \( L_{2} \) and RoFL with \( L_{\infty} \). Although our zero-knowledge proof is more complex compared to the proof used in RoFL, replacing the Bulletproof protocol with zk-SNARKs significantly reduces the proving and verification time, contributing to the overall efficiency of our approach. Our comparison results suggest that our scheme can handle more complex secure federated learning models without compromising time performance.

In conclusion, this comparison demonstrates that our scheme, leveraging partially homomorphic encryption and zero-knowledge proofs, achieves substantial improvements over other schemes in terms of efficiency and scalability.

\section{Discussion}
\label{sec:Discussion}

\subsection{Related Work}
The landscape of Secure Federated Learning (SFL) has seen significant advancements aimed at balancing data privacy, communication efficiency, and robustness against adversarial threats. One notable area of exploration involves the use of cryptographic techniques beyond commonly discussed approaches, such as multi-party computation (MPC) and differential privacy (DP). For instance, works~\cite{knott2021crypten} have demonstrated the potential of MPC in distributed learning scenarios, achieving strong privacy guarantees by securely computing updates without data exposure. However, these solutions often incur significant communication and computational overhead, limiting their scalability in real-world applications. Meanwhile, differential privacy approaches~\cite{pentyala2022training}, including differentially private stochastic gradient descent (DP-SGD), have focused on safeguarding individual data contributions during training. While these techniques effectively balance privacy and utility, they often struggle with increased noise and reduced model accuracy, prompting further research on optimizing this trade-off.

Another active area of research focuses on Byzantine-robust federated learning, ensuring the learning process remains resilient to malicious clients. Methods such as Byzantine-resilient stochastic aggregation~\cite{wu2023byzantine} and robust statistics~\cite{data2021byzantine} have been developed to mitigate adversarial behaviors, typically by reducing the influence of tampered or outlier updates. However, these methods are often designed for plaintext settings, and their adaptation to secure federated learning introduces challenges, especially when computations must remain encrypted.

Privacy-preserving federated optimization has also explored hybrid models, such as using secure enclaves (e.g., Intel SGX) to isolate and secure sensitive computations~\cite{cerdeira2020sok}. While these approaches offer strong security guarantees, they are heavily hardware-dependent, limiting their applicability in diverse deployment environments. Additionally, communication-efficient federated learning techniques, such as federated dropout~\cite{wen2022federated}, aim to reduce the communication burden by selectively pruning updates during training. However, these approaches fail to address the complexities introduced by secure aggregation in adversarial settings, leaving gaps in their robustness.

Our work contributes to this evolving landscape by exploring the gap between privacy, efficiency, and Byzantine robustness, proposing a novel framework that seeks to overcome the limitations highlighted in these studies.

\subsection{Limitation}
Similar to most current secure aggregation-based SFL frameworks~\cite{RoFL, ACORN, EIFFeL}, our approach does not maintain a fully encrypted state throughout the entire training process, unlike secure multi-party computation methods. Specifically, while secure aggregation ensures that intermediate client updates remain encrypted during transmission, the model parameters at each iteration are visible to both the clients and the encryption server, $\mathbb{S}_E$. This exposure creates a potential vulnerability, where adversaries might exploit these model parameters to infer sensitive information.

One notable risk arises when the number of clients is small, especially in the case of only two clients. In this scenario, a malicious client could potentially observe its own updates and infer the updates of the other client across iterations. This knowledge could be used to mount targeted attacks or infer private data from the learning process. However, as the number of clients increases, this attack becomes more difficult. The presence of multiple clients dilutes the influence of any single client's update, making it challenging to isolate and infer the contributions of specific participants.
While our framework introduces efficiency and practical privacy guarantees, this limitation underscores the trade-off between maintaining computational efficiency and achieving the highest level of security. Future work could explore hybrid approaches that balance secure aggregation and more comprehensive encryption techniques to further mitigate this risk.

\section{Conclusion}
We presented \ourmethod{}, an efficient framework for Byzantine-robust secure aggregation in Secure Federated Learning. By leveraging a Zero-Knowledge Proof protocol toolkit and a novel dual-server architecture, \ourmethod{} ensures robust aggregation while maintaining data privacy and achieving a 85x computational speedup compared to traditional approaches. Our work addresses critical gaps in existing SFL systems, offering a practical and scalable solution for data-sensitive applications.

\clearpage






%

\bibliographystyle{IEEEtran}  
\bibliography{ref}

\appendix
\label{sec:appendix}


\subsection{Notation}

\vspace{-10pt}

\label{sec:Notation}
\begin{table}[!h]
\setlength{\tabcolsep}{1.0mm}{
\footnotesize
\centering
\begin{tabular}{|c|l|}
\hline
\textbf{Notation} & \textbf{Definition} \\ \hline
$m$ & Number of parties (clients) \\ \hline
$\beta_0$ & Initial model parameters \\ \hline
$\mathbf{g}^*$ & Reference update computed by $\mathbb{S}_C$ using its validation data \\ \hline
$\mathbf{g}_i$ & Local update of the $i^{th}$ client \\ \hline
$\alpha_i$ & Aggregation weight of the $i^{th}$ client \\ \hline
$\mathcal{C} (\hat{\mathbf{g}_i})$ & Encrypted local gradient uploaded by the $i^{th}$ client \\ \hline
$\mathcal{C} (\hat\beta)$ & Encrypted final global model parameter \\ \hline
$\mathcal{E}(\cdot)$ & Encryption function of homomorphic encryption scheme \\ \hline
$\mathcal{D}(\cdot)$ & Decryption function of homomorphic encryption scheme \\ \hline
$\hat\beta$ & Aggregated global model parameter, $\hat\beta = \frac{\sum_{i=1}^m \hat\beta_i}{m}$ \\ \hline
$\eta$ & Optimization learning rate \\ \hline
\end{tabular}
\caption{Summary of Notations and Definitions}
\label{tab:notations}
}
\end{table}
\vspace{-10pt}

\subsection{Baseline Protocol with FHE}
\vspace{-5pt}

\label{sec:fhe_protocol}
The detailed protocol design with fully homomorphic encryption is presented in \autoref{fig:fhe_protocol}.
The CKKS scheme is configured with the following parameters:

\vspace{2pt}\noindent$\bullet$ \textbf{Polynomial modulus degree} as $2^{16}$, corresponding to a polynomial ring of degree $32768$, which balances performance and security.

\vspace{2pt}\noindent$\bullet$ \textbf{Scaling factor} as $2^{30}$, allowing high precision for encrypted real number arithmetic by scaling plaintext values.

\vspace{2pt}\noindent$\bullet$ \textbf{Coefficient modulus sizes} as $\{60, 30, 30, 30, 60\}$ bits. This chain of moduli enables multiple homomorphic operations while maintaining the noise budget.

The standard CKKS protocol does not allow range comparison. Our protocol requires a maximum function, so we use a min-max approximation \cite{lee2021minimax}, as we only need to reject negative values rather than achieve a full comparison.

\begin{figure}[!h]
\small
\centering
\fbox{
\parbox{.95\linewidth}{
\textbf{Setup:}
\begin{enumerate}[itemsep=0pt]
    \item Server $\mathbb{S}_C$ sends initial model $\beta_0$, loss function $\ell$, optimizer $\mathcal{O}$ and learning rate $\eta$ to each client.
    \item Server $\mathbb{S}_E$ sends \texttt{ek} to each client 
\end{enumerate}
}}
\fbox{
\parbox{.95\linewidth}{
\textbf{$i^{\text{th}}$ Client:}
\begin{enumerate}[itemsep=0pt]
    \item Compute $\mathbf{g}_i$ by ~\autoref{eq:beta_i}
    \item Encrypt $\mathcal{C}(\mathbf{g}_i) = \mathcal{E}(\mathbf{g}_i, \texttt{ek})$
    
    
    
    
    \item Send $\mathcal{C}(\mathbf{g}_i)$ to server $\mathbb{S}_C$
\end{enumerate}
}}
\fbox{
\parbox{.95\linewidth}{
\textbf{Server $\mathbb{S}_C$}
\begin{enumerate}[itemsep=0pt]
    \item Compute $\mathcal{C}(\text{TS}_i)$ and $\mathcal{C}(\Tilde{\text{TS}}_i)$ for each client by ~\autoref{eq:normalizd fltrust score} with encrypted $\mathcal{C}(\mathbf{g}_i)$.

    \item Compute $\mathcal{C}(\mathcal{H}) = \sum_{i \in \text{All clients}}\mathcal{C}(\Tilde{\text{TS}}_i)\cdot \mathcal{C}(\mathbf{g}_i)$    
    \item Compute new global model $\mathcal{C}(\beta_{\text{new}}) = \frac{\mathcal{C}(\mathcal{H})}{\mathcal{C}(\text{TS})}$
    \item Send $\mathcal{C}(\beta_{\text{new}})$ to clients
\end{enumerate}
}}
\fbox{
\parbox{.95\linewidth}{
\textbf{$i^{\text{th}}$ Client:}
\begin{enumerate}[itemsep=0pt]
    \item Decrypt $\mathcal{C}(\beta_{\text{new}})$
    \item Send the updated global model $\beta_{\text{new}}$ to each client for next round 
\end{enumerate}
}}

\caption{Protocol with fully homomorphic encryption}
\label{fig:fhe_protocol}
\end{figure}

\end{document}